\documentclass[myepj-spec,pdftex]{mySvjour}
\usepackage{graphicx}
\usepackage{hyperref}
\usepackage{color}
\usepackage{xspace}
\usepackage[backend=bibtex,hyperref=true,block=ragged,sorting=none,maxbibnames=5,style=numeric-comp]{biblatex}
\usepackage{verbatim}
\usepackage{amsmath,amssymb,amsfonts} % Typical maths resource packages
\usepackage{tabularx}
\usepackage[switch, modulo]{lineno}
\usepackage{booktabs}
\usepackage{makecell}
\usepackage[detect-all]{siunitx}
\DeclareSIUnit{\radiationlength}{$X_0$}
\DeclareSIUnit\barn{b}

\usepackage{lmodern,bm}
\usepackage[T1]{sansmath} 
\SetMathAlphabet{\mathsfbf}{sans}{\sansmathencoding}{\sfdefault}{bx}{sl}
\usepackage{etoolbox}
\AtBeginEnvironment{sansmath}{}{}{}

\usepackage{lipsum}

\definecolor{darkblue1}{rgb}{0,0,.2}
\definecolor{darkblue}{rgb}{0,0,.2}
\definecolor{darkred}{rgb}{0.5,0,0}
\pagecolor{white} % Background color
\color{black}     % Text color
\hypersetup{breaklinks=true, 
            colorlinks=true, 
            linkcolor=darkblue1, 
            menucolor=darkblue1, 
            urlcolor=darkblue1,
            citecolor=darkblue1,
            pdftitle={},
            pdfauthor={},
            pdfsubject={},
            pdfkeywords={},
            pdfproducer={}
}

\bibstyle{plain}
\newcommand{\bi}{\begin{itemize}}
\newcommand{\ei}{\end{itemize}}

\newcommand{\ben}{\begin{enumerate}}
\newcommand{\een}{\end{enumerate}} 

\newcommand{\bt}[1]{\begin{table}[tb]\begin{tabular}{#1} \hline\hline  \\[-1.0em]}
\newcommand{\et}[2]{\hline\hline \end{tabular} \caption{#1} \label{#2} \end{table}}

\newcommand{\be}{\begin{equation}}
\newcommand{\ee}{\end{equation}}

\newcommand{\bea}{\begin{eqnarray}}
\newcommand{\eea}{\end{eqnarray}}

\newcommand{\bc}{\begin{comment}}
\newcommand{\ec}{\end{comment}}

\newcommand{\mev}{\ensuremath{\mathrm{\,Me\kern -0.1em V}}\xspace}
\newcommand{\gev}{\ensuremath{\mathrm{\,Ge\kern -0.1em V}}\xspace}

\begin{document}

\twocolumn[{%
  \begin{@twocolumnfalse}

    \begin{flushright}
      \normalsize
%      \today
    \end{flushright}

    \vspace{-2cm}

\title{\Large\boldmath Detection of collinear high energetic di-photon signatures with Micromegas Detectors}
%
% \affiliation command applies to all authors since the last
% \affiliation command. The \affiliation command should follow the
% other information
% \affiliation can be followed by \email, \homepage, \thanks as well.

\author{Friedemann Neuhaus\inst{1} \and Elisa Ruiz Choliz \inst{1} \and Matthias Schott \inst{1}}
\institute{\inst{1} Institute of Physics, Johannes Gutenberg University, Mainz, Germany}

% \noaffiliation

\abstract{
The search for weakly interacting, light particles that couple to photons received significant attention in recent years. When those particles are produced at high energies, they lead to two, nearly collinear photons after their decay and hence can be detected by an electromagnetic calorimeter system. The typical dominant background in searches for those high energetic weakly particles are single, high energetic photons, which leave similar signatures in a standard calorimeter system. One promising approach to separate signal from background events is to  employ a dedicated pre-shower detector in front of the calorimeter that can distinguish one- and two-photon signatures. In this work we present a conceptual design of such detector which is able to separate one from two collinear photon signatures with efficiencies between \SI{20}{\percent} to \SI{80}{\percent} for two photons separated by \SI{100}{\micro\meter} to \SI{2000}{\micro\meter}, respectively, with energies above \SI{300}{\GeV} and a background rejection of more than \SI{90}{\percent}. Our pre-shower detector design has an active surface area of \SI[product-units=single]{10 x 10}{\centi\meter\squared}, a depth of \SI{230}{\milli\meter} and is based on Micromegas detectors, thus offering a cost effective solution.}

\maketitle
  \end{@twocolumnfalse}
}]

\tableofcontents

\section{Introduction}

Axions and other very light axion-like particles (ALPs) \cite{Axions1,Axions2} appear in many extensions of the Standard Model and are well motivated theoretically: ALPs can solve the well-known strong CP problem \cite{CPProblem1,CPProblem2}, act as a dark matter candidate \cite{AxionsDM} and could also explain the famous muon $(g_\mu-2)$ discrepancy~\cite{Bauer:2017nlg}. The experimental effort to search for ALPs as dark matter candidates is ongoing and has been considerably intensified in recent years, leading to the proposal and construction of a wide range of dedicated experiments. ALPs with masses in the \si{\MeV} regime and above are typically searched for in collider experiments, such as the ForwArd Search ExpeRiment (FASER)~\cite{faserOverview,faserPhysicsReach}, which was installed in the LHC in March 2021 and is situated \SI{480}{\meter} along the line-of-sight of the proton collisions in front of the ATLAS interaction point at the LHC. ALPs are produced in collider experiments with typical momenta that are much larger than their mass, hence the generic experimental signatures are two highly collinear photons. While standard calorimeters can measure the total energy of those photons, they have no means to distinguish their electromagnetic showers if the distance between them is smaller than the cell-size of the calorimeter system. 

In this paper, we present a conceptual pre-shower detector design that is able to separate high energetic one-photon events from collinear two-photon events. This separation is achieved by alternating layers of tungsten, that act as passive material triggering photon conversions and subsequent electromagnetic showers; and Micromegas detectors, for the reconstruction of charged particles. In total three layers of tungsten and three Micromegas detectors with an active area of \SI[product-units=single]{10 x 10}{\centi\meter\squared} and a two-dimensional readout are used. A simulation based on \textsc{Geant4} was developed to evaluate the expected performance of the preshower detector, using deep neural networks for the event classification. The simulation was validated by test beam measurements at the MAMI (MAinz MIcrotron) accelerator. The optimisation of the detector layer as well as the prototype detector which was used during the test beam measurements is discussed in \autoref{sec:detector-layout}. The simulation as well as its validation is summarized in \autoref{sec:simulation}, followed by a discussion of the photon identification algorithm (\autoref{sec:identification}) and the results (\autoref{sec:results}).

\section{Detector Layout and Optimization}\label{sec:detector-layout}

\begin{figure*}[tb]
\centering
\hfill\includegraphics[width=0.56\textwidth]{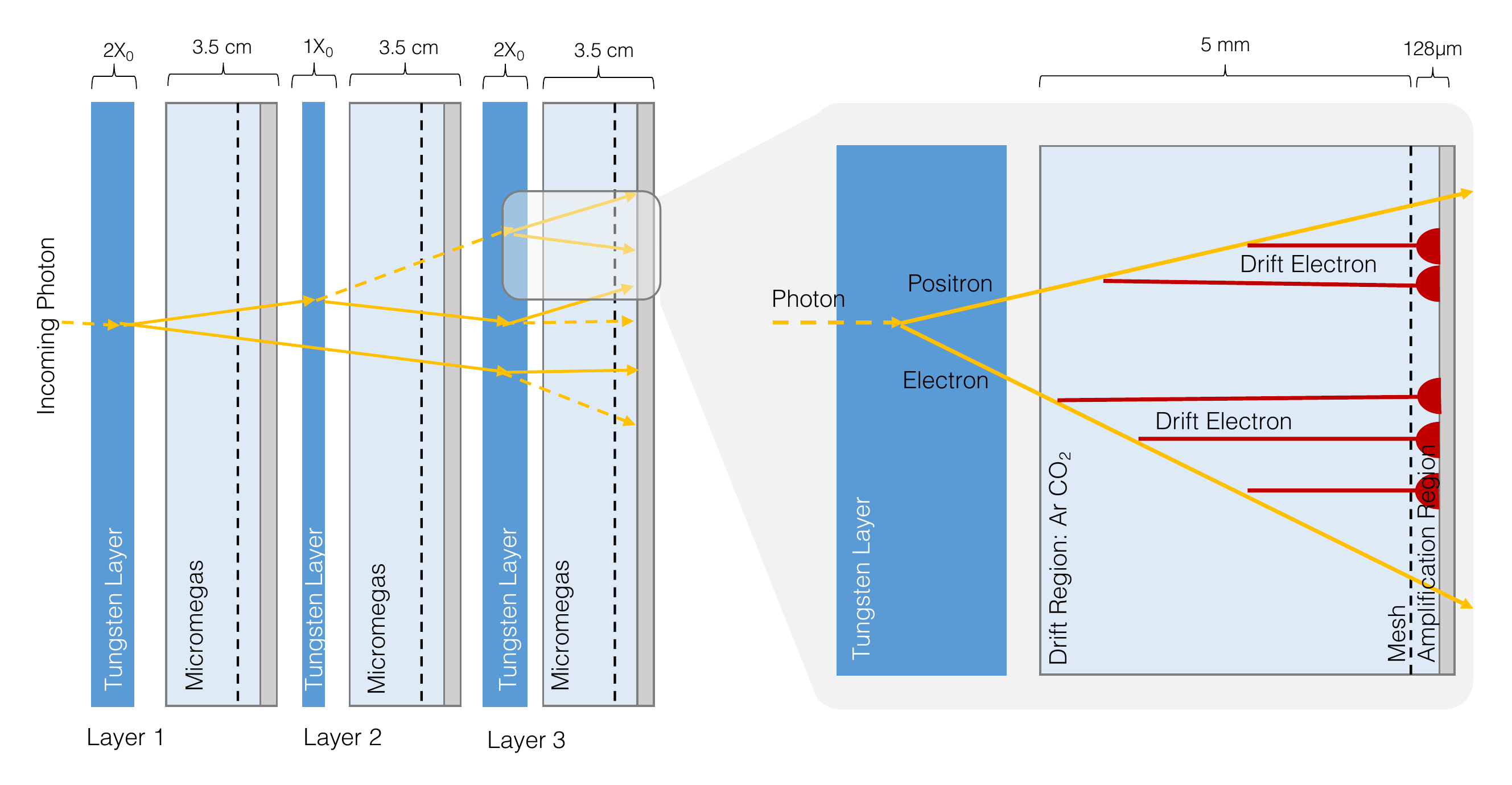}\hfill
\includegraphics[width=0.40\textwidth]{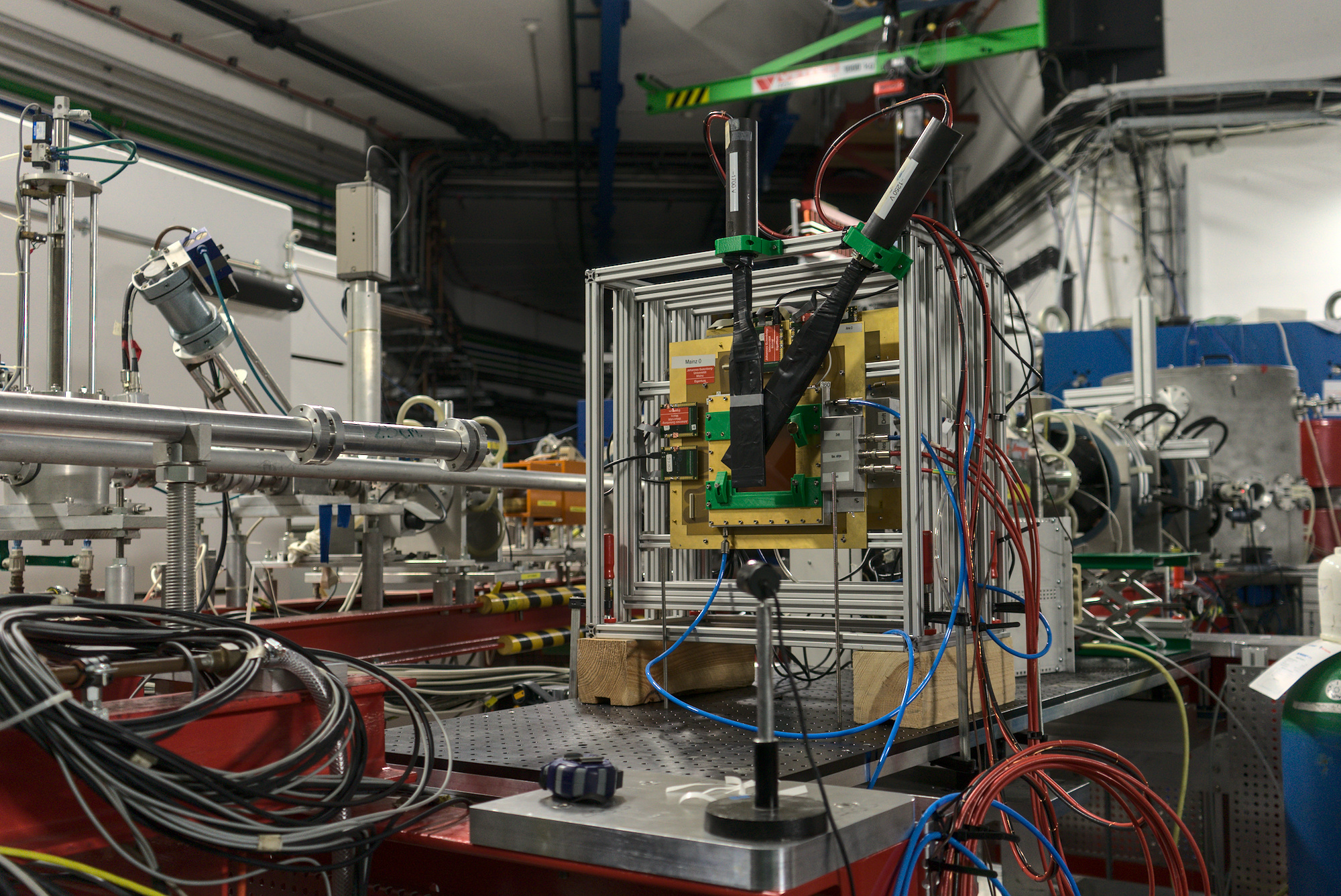}
\caption{Schematic layout of the detector setup (left) as well as picture of the prototype in the test beam setup (right).}
\label{fig:Detector}
\end{figure*}

The schematic design of the pre-shower detector is shown in \autoref{fig:Detector}. The number of passive tungsten layers as well as the number of active Micromegas detectors was fixed in our design approach as our total pre-shower detector should not exceed a potential minimal depth of $\approx$20 cm, however, the thickness of the tungsten layers was subject to optimization. 

We propose to use standard Micromegas detectors with a two-dimensional readout, which have been originally designed by the MAMMA collaboration~\cite{Byszewski:2012zz} and are described in detail in \cite{Lin:2014jxa}. They have an active area of \SI{10 x 10}{\centi\meter} and a depth of \SI{3.5}{\centi\meter} and have proven to operate with a high reliability, a very good spatial resolution of $\SI[parse-units=false]{<200}{\micro\meter}$ and also, they are not subject to dead-times from sparks. The drift-region, filled with a gas mixture of \SI{93}{\percent} Argon and \SI{7}{\percent} CO$_2$, has a depth of \SI{5}{\milli\meter}, while the amplification region has a depth of \SI{128}{\micro\meter}. The readout electrode of each detector comprises 360 copper readout strips in two separate, orthogonal layers with a strip pitch of \SI{250}{\micro\meter}, allowing to measure two-dimensional spatial information. The readout strips of the upper layer (defined as $y$-layer) are printed directly on top of the PCB and are covered by the resistive strips with a resistivity of \SI{20}{\mega\ohm\per\centi\meter}. The lower layer (defined as $x$-layer) is separated from the upper layer by \SI{70}{\micro\meter} of FR4, i.e. the same material used as isolating material in the PCB. The readout strips of the $x$-layer have a width of \SI{200}{\micro\meter} and are placed parallel to the resistive strips, while the strips in the $y$-layer have a width of \SI{80}{\micro\meter} and are perpendicularly placed with respect to the resistive strips. The larger width of the $x$-layer readout strips compensates for their weaker capacitive coupling. The woven mesh is made of stainless steel with a density of \SI{157}{lines\per\centi\meter} and a diameter of \SI{18}{\micro\meter}. It is mounted on support pillars of \SI{0.4}{\milli\meter} diameter and covers an area of \SI{10 x 10}{\centi\meter} which defines the active area of the detector. The support pillars are placed along a regular matrix with \SI{2.5}{\milli\meter} spacing in both directions.

The data acquisition is based on the RD51 Scalable Readout System (SRS)~\cite{Martoiu:2013aca, Martoiu:2011zja}. The signal processing of the detector is based on the Analog Pipeline Voltage chip with \SI{0.25}{\micro\meter} CMOS technology (APV25)~\cite{APVJones} where the analog signal data is transmitted via HDMI cables to SRS electronics. The SRS electronics process the analog signal which is then further analyzed. It should be noted that the height of the recorded readout signal is related to the charge induced on the readout strip\footnote{The APV25 integrates the induced current for \SI{75}{\nano\second}, so the resulting charge is only part of the total induced charge.}. The photon identification algorithms are based on the measured integrated charge in each strip of both layers in all three Micromegas detectors, since the measured charge is correlated with the number of created charged particles in the drift region of the detector. No timing information was used. 

The figure of merit for the optimization of the system was the correct identification of two photon events with a separation of \SIrange{200}{2000}{\micro\meter} and energies of \SIrange{100}{3500}{\GeV} and one-photon events with energies of \SIrange{200}{7000}{\GeV}. Several layouts have been tested, using the simulation and classification algorithms described in more detail in Sections \ref{sec:simulation} and \ref{sec:identification}. Optimal results have been achieved for \SI{2}{\radiationlength} of tungsten in front of the first and the last active layer and \SI{1}{\radiationlength} in front of the second active layer, with a separation of \SI{7}{\centi\meter} between the active planes.

\begin{figure*}[tb]
\centering
\includegraphics[width=0.49\textwidth]{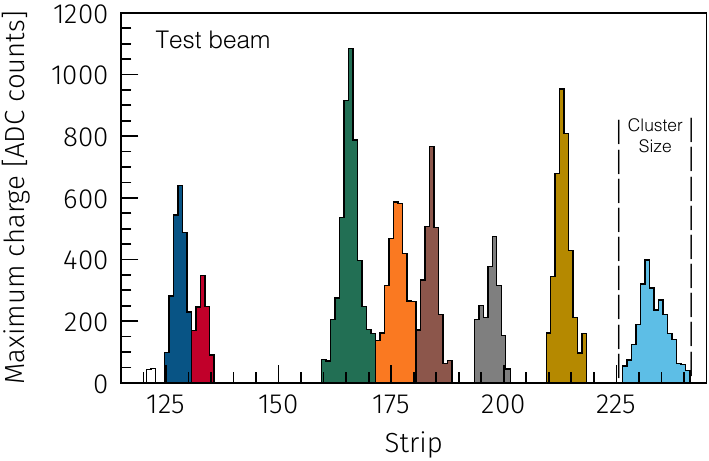}\hfill
\includegraphics[width=0.49\textwidth]{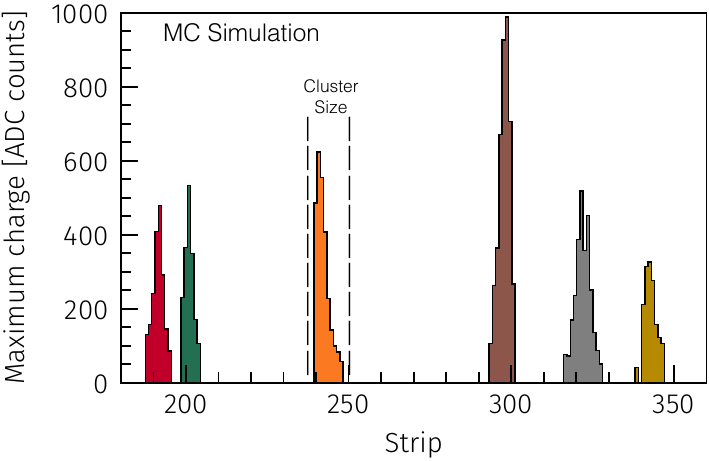}
\caption{Single event in the Y-layer of the middle detector from the test beam (left) and from simulation (right). The different colors indicate the selected clusters.}
\label{fig:clusters}
\end{figure*}

\section{Prototype Detector Simulation and Validation with test beam Measurements}\label{sec:simulation}

Since it is highly challenging to organize test beam measurement campaigns which would allow to measure directly the pre-shower detector performance for the separation of collinear photon signatures at energies above several hundred \si{\GeV}, an alternative approach has been taken: one specific detector layout was setup and studied in test beam measurements. These measurements are then used to validate the \textsc{Geant4} based simulation, which is used in a second step to estimate the detector performance for the separation of collinear photon signatures. 

Incident electrons and photons may create an electromagnetic shower via Bremsstrahlung and pair-production in the tungsten layers, leading to electron positron pairs which enter the drift-region of the Micromegas detector, where they will interact with and ionize gas atoms. The resulting ionization electrons will drift to the amplification region and hence, induce a signal at the corresponding readout strips. Typically, signals due to one incident primary particle, i.e. electron or photon, will induce electric signal in neighbouring readout strips, leading to clusters. To find clusters in a first step, peaks are searched in the maximum charge distribution. The extent of the cluster is determined iterating the neighbouring strips of each peak in both directions until either more than two strips without charge are found or the charge is increasing.
This allows to separate close-by clusters as long as they are not fully overlapping. The size of those cluster is then defined by the number of strips hit. Examples of the recorded charge distribution per readout strip for one electron event that produced a shower in the tungsten layers are shown in Figure \ref{fig:clusters} for the test beam measurement as well as the simulation, where several clusters can be clearly identified. 

The recorded signal height as well as its time evolvement within one cluster depends on one hand, on the number of corresponding drift electrons and on the other hand, on the applied amplification voltage. The position of one cluster is defined as the weighted mean of all associated readout-strips with their measured signal. For the photon identification algorithm the maximum charge from each strip is taken and fed as an input to the neural network. It is therefore important that the number of clusters as well as their charge distribution is correctly described in the simulation. 

The test beam setup of the pre-shower detector consists also of three active Micromegas planes that are interleaved by three layers of tungsten with a variable thickness of \SI{1}{\milli\meter} to \SI{6}{\milli\meter} (\autoref{fig:Detector}). The distance between the passive tungsten and active Micromegas layers was \SI{4}{\milli\meter} with \SI{75}{\milli\meter} distance between the Micromegas layers to allow accessing the tungsten layers, yielding a total length of the prototype detector of \SI{230}{\milli\meter}. Different thicknesses of the tungsten layers have been chosen during the test beam measurements in order to validate the detector simulation over a larger parameter space. The test beam measurements have been conducted in 2020 at the MAMI accelerator facility at the Johannes Gutenberg University Mainz. MAMI provides a quasi-continuous electron beam with energy up to \SI{1.5}{\GeV}. A beam energy of \SI{855.1}{\MeV} was used for the measurements presented in this paper. Clean signals have been recorded by applying drift and amplification voltages of $V_D = \SIlist[list-final-separator={, }]{300;200;225}{\volt}$ and $V_A = \SIlist[list-final-separator={, }]{560;565;570}{\volt}$ respectively for the three Micromegas detectors. The beam axis was chosen to be perpendicular to the readout-panels. 

The test beam setup was then implemented also in the \textsc{Geant4} simulation framework. The interaction of the incident electrons with the detector material and the propagation of the particles of the resulting electromagnetic shower through the detector, in particular the energy deposits within the drift volume of the Micromegas detectors, are simulated directly by \textsc{Geant4}. The position of each energy deposit is stored. Then the charge deposits are added to the strips based on event shapes extracted from data. The amplitude is determined by multiplying the deposited energy from \textsc{Geant4} with a fixed gain factor for each layer. This allowed getting a reasonable description of the maximum charges for a first study without requiring a detailed simulation of the gas transport, amplification and digitization. Once the simulation has been implemented, the detector response of \num{100000} events of incoming electrons with a energy of \SI{855.1}{\MeV} has been simulated.

\begin{figure*}[tb]
\centering\includegraphics[width=\textwidth]{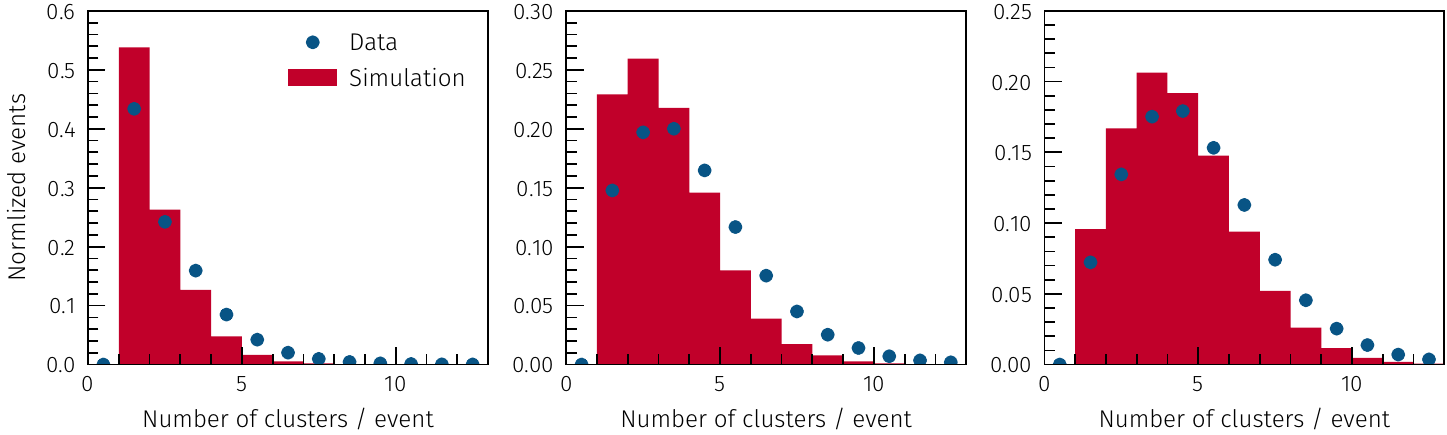}\\
\includegraphics[width=\textwidth]{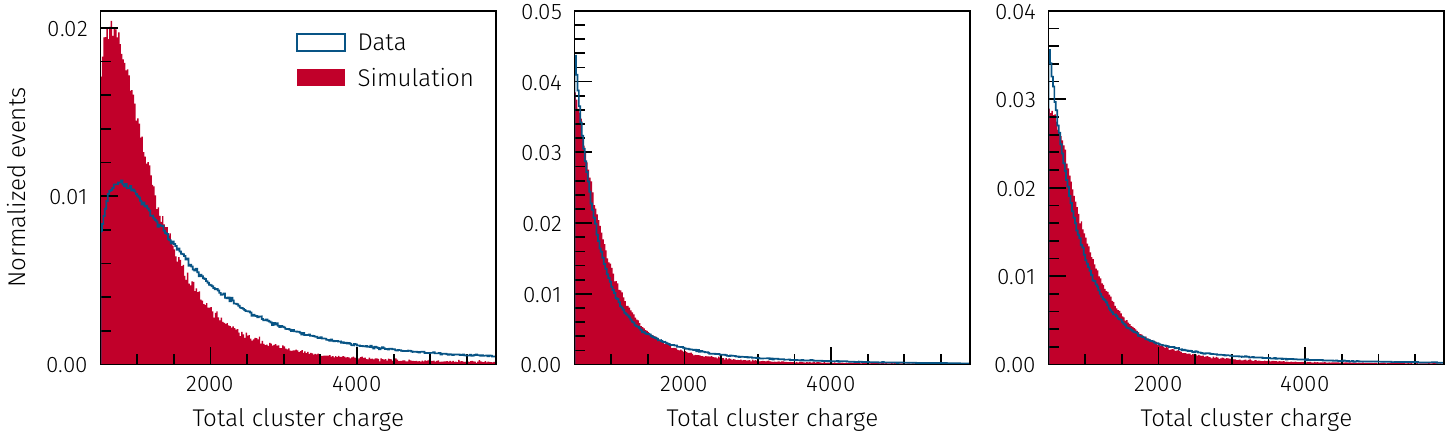}
\caption{Comparison of the number of reconstructed clusters (top) and the total charge per cluster (bottom) in the $y$-layers of all detectors in simulation (bars) and test beam measurements (points/line).}
\label{fig:Simulation}
\end{figure*}

In a second step, the cluster-finding algorithm has been applied to the test beam data and the simulation. The electronic response parameters have been tuned such that they describe the observed cluster properties, in particular the cluster-size as well as the signal height. It has been tested for all available setups of the prototype detector used during the test beam campaign.
\autoref{fig:Simulation} shows the number of reconstructed clusters as well as the total charge per cluster in all detector layers in simulation and measurement for the setup with \SI{1}{\radiationlength} in front of the first two layers and \SI{2}{\radiationlength} in front of the last layer. The mean of the number of clusters between data and simulation agrees at a \SI{10}{\percent} level, however, systematically more clusters are reconstructed in data than simulation.
This can be explained by dead channels not being explicitly modelled as well as by differences in the detector gain.
Similarly, the total cluster charge is larger in data than in simulation. The largest deviation is found in the first detector layer which can be explained by sub-optimal bias voltage settings during the test beam.
The effect of this disagreement on the final photon identification algorithms has been tested by scaling the simulated charge deposits. 

Once the simulation was tuned and validated for the electron beam, the same setup was used to simulate single and di-photon events with various event kinematics: events are evenly distributed in the parameter space with the photon energies within \SIrange{200}{7000}{\GeV} for single photons and \SIrange{100}{3500}{\GeV} for di-photon events. The separation between the two photons is taken to be within \SIrange{0}{2000}{\micro\meter}. In total 6 million single photon and 12 million di-photon events are simulated. The higher number of di-photon events is chosen to compensate for the lower event yield when applying filter criteria on the photon conversion.

\section{Di-photon Identification}\label{sec:identification}

Single photon events are expected to produce fewer clusters in each layer compared to di-photon events. Moreover, the cluster distribution is expected to be narrower for the former, compared to the latter. For an optimal classification of single- and di-photon events, a deep convolutional neural network based classifier has been used. 

For the nominal geometry with \SI{1}{\radiationlength} worth of absorber material in front of the first detector, \SI{18.7}{\percent} of the photons produce less than \num{25} hits.
For the alternative geometry with \SI{2}{\radiationlength} of absorber, this number drops to only \SI{10}{\percent} of photons producing less than \num{25} hits due to the photon shower being initiated earlier. \textit{To ensure that the training consists mostly of events where the photons converted at least \num{25} hits per photon are required across the three detectors.} The training is performed using Keras~\cite{keras} with Tensorflow~\cite{tensorflow} as the back-end.

The network takes inputs in two stages. First, the charge on each strip (total of 2160 inputs) is fed into one dimensional convolutions. The output of the convolution is reduced to a single dimension by a flatten layer. Afterwards, the total energy of the event is inserted as an additional input and passed into three dense layers. Here, the total event energy is smeared to reflect the resolution of a typical electromagnetic calorimeter. Finally, a layer with only a single output neuron is used. The activation function is a rectified linear unit for all layers except for the output layer which uses a sigmoid as the activation to ensure an output between \num{0} and \num{1}. The labels are chosen such that the output is \num{0} for single photon events and \num{1} for di-photon events.

The network architecture has been optimized by minimizing the loss defined by binary cross-entropy. The parameters for the convolution layers, number of fully connected layers and number of neurons were determined by running a hyper-parameter scan where the architecture with the highest validation accuracy was selected. The parameters of the full network architecture are summarized in \autoref{tab:network-architecture}.

\begin{table}[tb]
    \centering
    \begin{tabular*}{\columnwidth}{@{\extracolsep{\fill} }lcccccr}
      \toprule
    Layer type      & Neurons & \makecell{Kernel size\\ / Filters} & Parameters \\ 
    \midrule
    Input (charges) &    &         &  \\
    1D Conv.  &    & \num{14} / \num{20} & \num{6800}  \\
    Max. pooling & &                      & \\
    Flatten         &    &         &       \\
    \midrule
    Input (energy)  &     &         &         \\
    Concatenate     &     &         &         \\
    Dense           & \num{40} &         & \num{576080} \\
    Dense           & \num{20} &         & \num{820} \\
    Dense           & \num{10} &         & \num{210} \\
    Output          & \num{1}   &         & \num{11}      \\
    \midrule
    Total           &     &         & \num{583921} \\
    \bottomrule
    \end{tabular*}
    \caption{Architecture of the neural network used for discriminating between single and di-photon events. Between each two layers, a dropout layer with a dropout rate of \num{0.1} is present. ReLu is used as the activation function for all layers except the output which uses a sigmoid function.}
    \label{tab:network-architecture}
\end{table}

\section{Results}\label{sec:results}

The expected performance of our preshower detector and di-photon reconstruction algorithm has been tested by using simulated events, which have not been used during the training of the network. First, we discuss the expected performance for several chosen single- and di-photon kinematics, followed by a discussion of the performance for ALP decays, where a convolution of several effects plays a role. 

\begin{figure*}[th]
\centering
\includegraphics[width=0.49\textwidth]{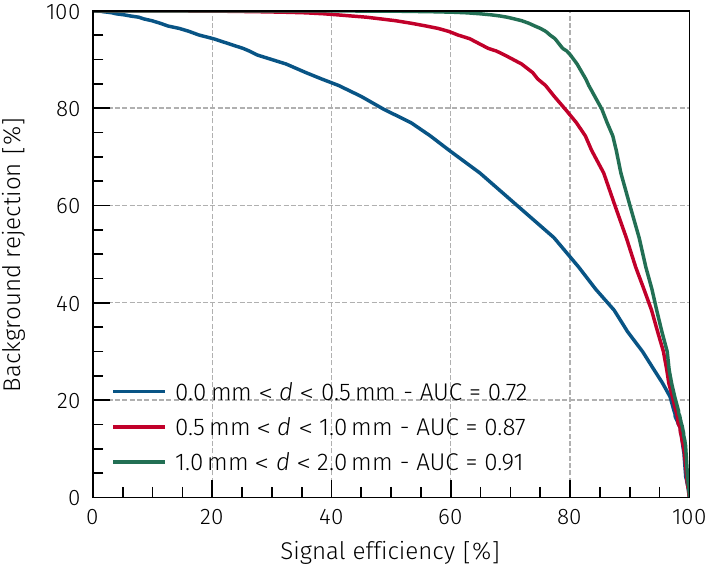}\hfill
\includegraphics[width=0.49\textwidth]{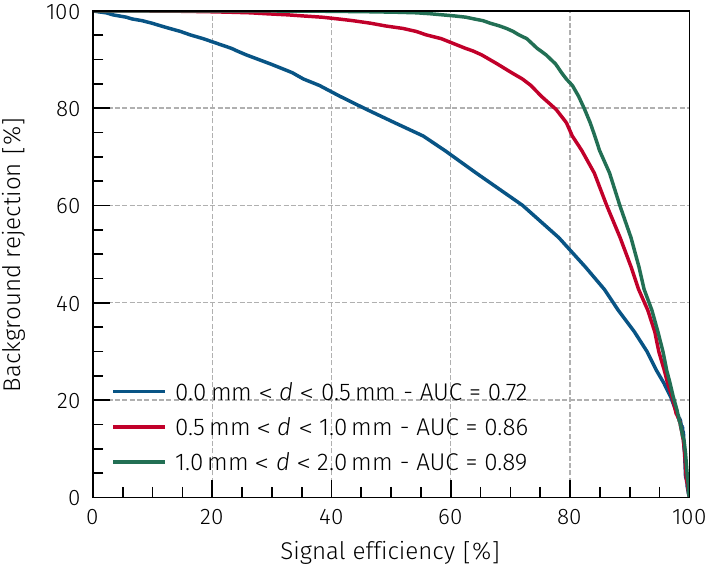}
\caption{ROC curves of di-photon events evaluated for different photon distance ranges for photons with equal energies of \SI{500}{\GeV} (left) and asymmetric energies of \SI{800}{\GeV} and \SI{200}{\GeV} (right); and \SI{1}{\TeV} for single photon backgrounds. AUC stands for \textit{area under the curve}.}
\label{fig:roc-curves}
\end{figure*}

\begin{figure*}[tb]
\centering
\includegraphics[width=0.49\textwidth]{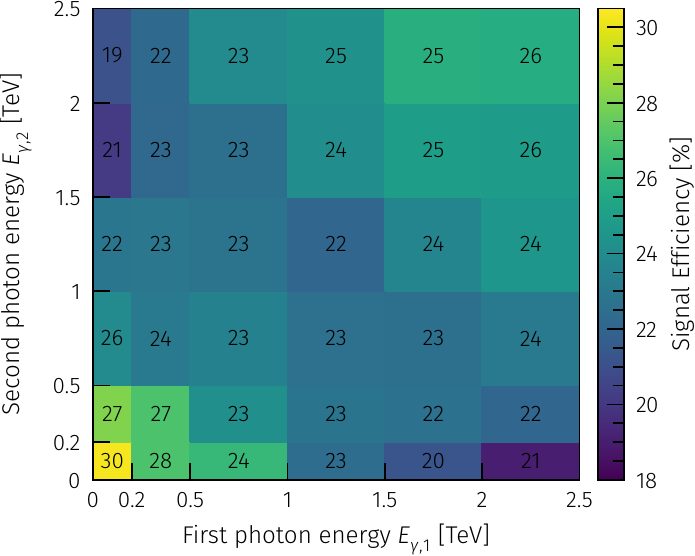}\hfill
\includegraphics[width=0.49\textwidth]{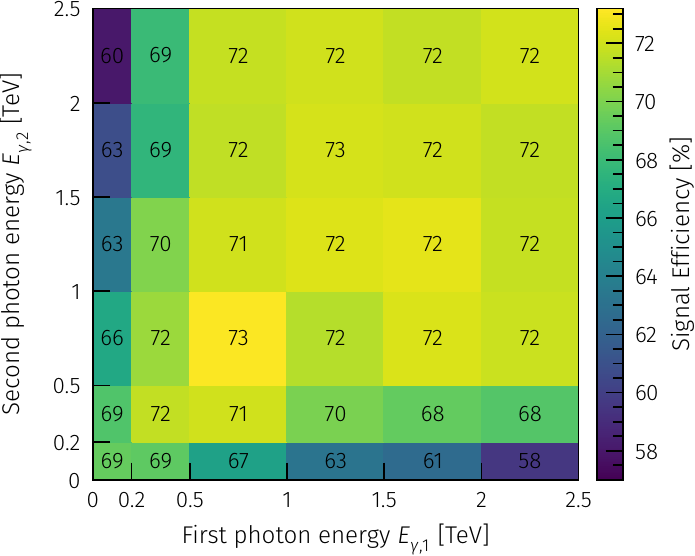}
\caption{Signal efficiency for di-photon events evaluated in dependence of the two photons energy in multiple bins for di-photon events with a separation of \SI[parse-numbers=false]{<300}{\micro\meter} (left) and for $\SI{500}{\micro\meter} < d < \SI{1000}{\micro\meter}$ (right) at an average background rejection of \SI{90}{\percent}.}
\label{fig:signal-efficiencies}
\end{figure*}

\begin{figure*}[tb]
\centering
\includegraphics[width=0.49\textwidth]{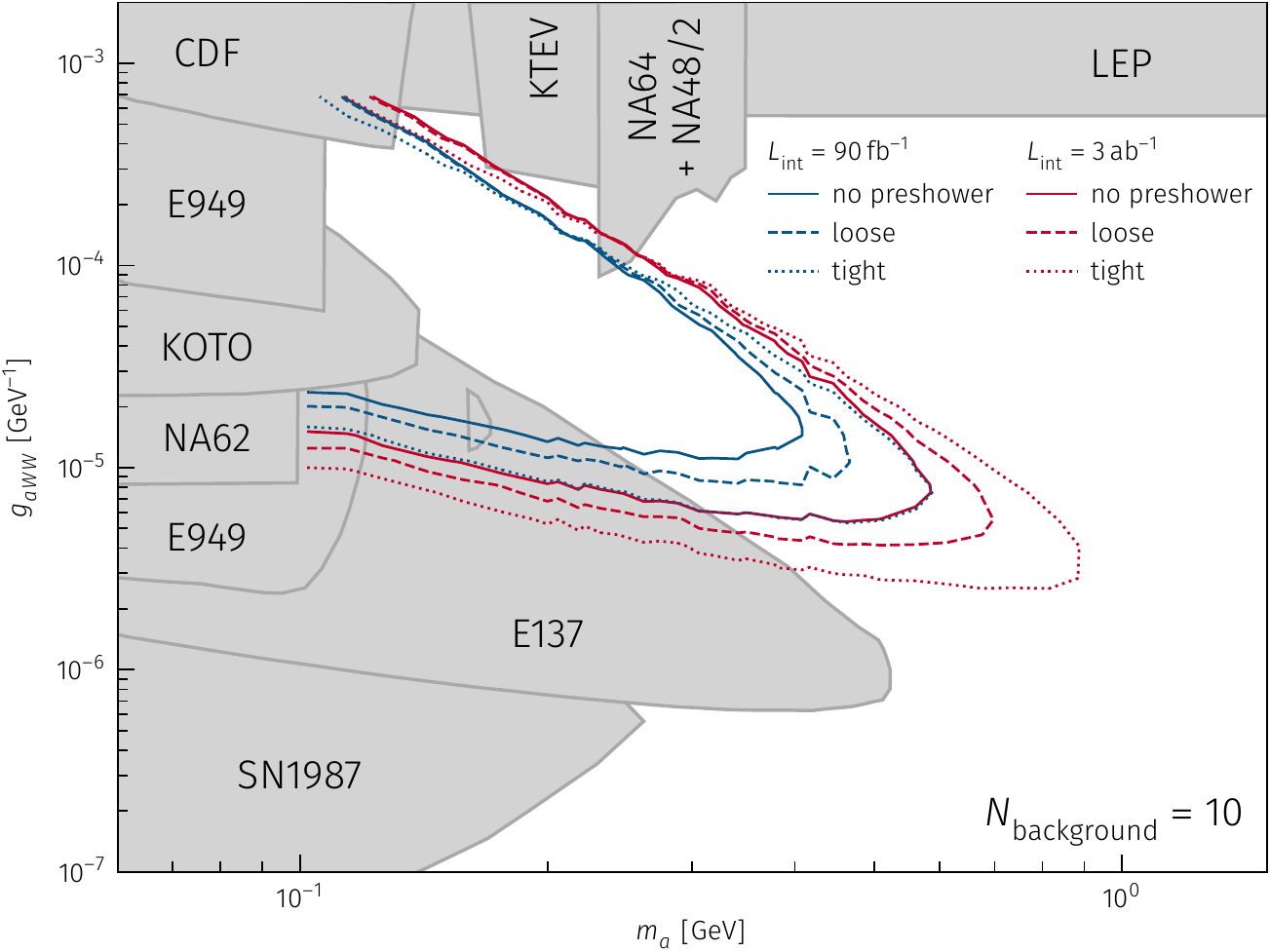}\hfill
\includegraphics[width=0.49\textwidth]{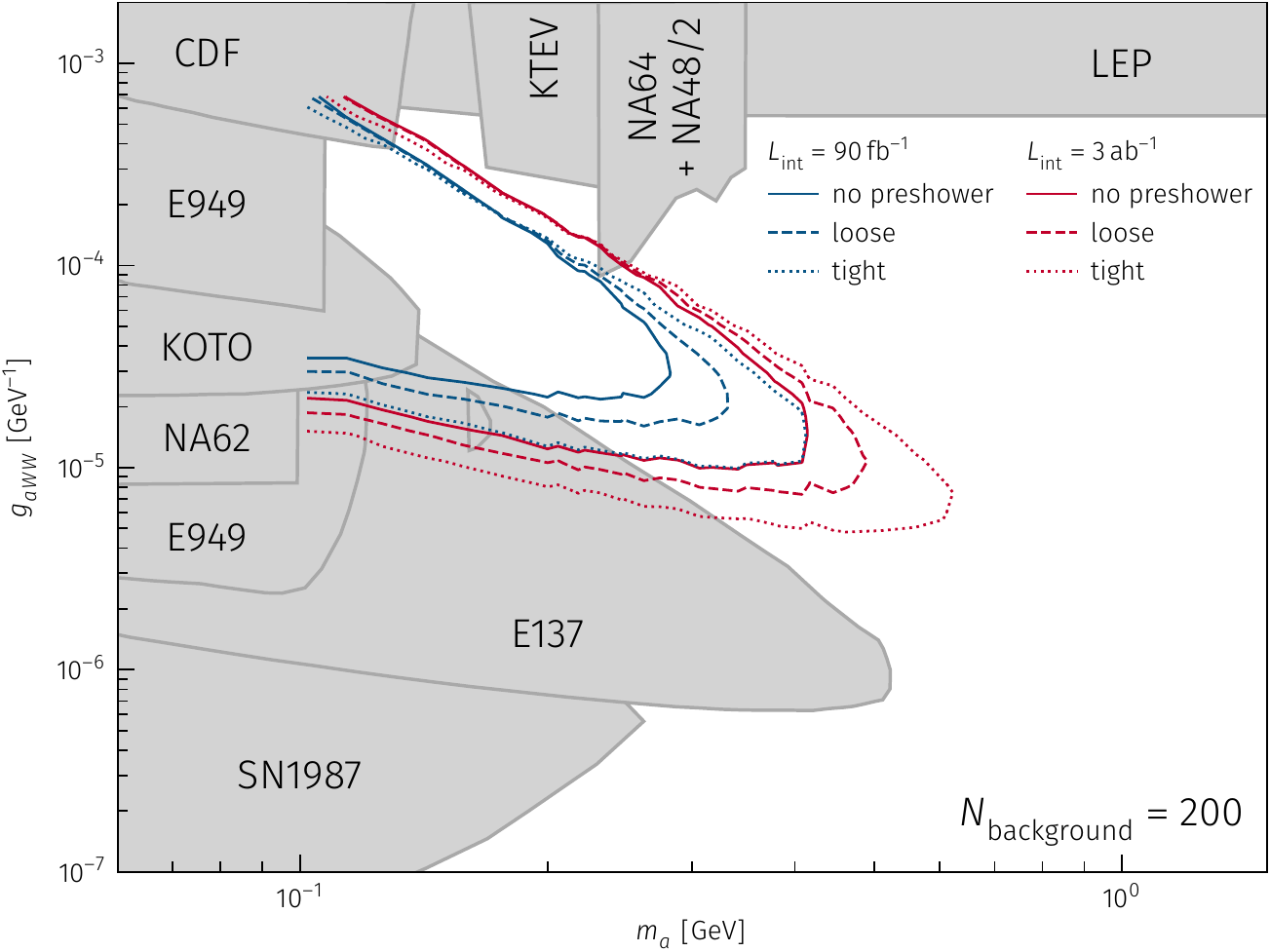}
\caption{Maximum limits on the $3\sigma$ sensitive region in the ALP mass and coupling plane for integrated luminosities of \SI{90}{\per\femto\barn} (blue) and \SI{3}{\per\atto\barn} (red) for the assumption of 10 (left) and 200 (right) background events.
    Three different scenarios are shown: no preshower (solid), the loose working point with \SI{90}{\percent} background rejection (dashed) and the thight working point with \SI{99.5}{\percent} background rejection (dotted).
    As the expected number of events is only available for a minimum separation of
    \SI{200}{\micro\meter} the reach is only calculated for this fiducial region.
    Already excluded regions are shown in grey and are taken from~\cite{klingParameterSpace}.}
\label{fig:physics-reach}
\end{figure*}

A good measure for the binary classifier is the Receiver Operating Characteristic (ROC) curve, defined by varying the classification-threshold and calculating the corresponding signal efficiency and the background rejection rate. An optimal classifier would yield an integral of the ROC curve of 1, while a pure random classifier for equal signal and background data-sets yields a value of 0.5.
\autoref{fig:roc-curves} shows the ROC curves for two photon signatures with energies of \SI{500}{\GeV} and distances between them of \SIrange{0}{500}{\micro\meter}, \SIrange{500}{1000}{\micro\meter} and \SIrange{1000}{2000}{\micro\meter} (signal) and single photon events with energies of \SI{1}{\TeV} (background).
Figure \ref{fig:roc-curves} also shows the results for the same samples, however, the energies of the di-photon signals are now not equally shared but \SI{800}{\GeV} and \SI{200}{\GeV}.
We observe only minor differences between the symmetric and asymmetric energy split, however, as seen in the area under the curve (AUC) the efficiency for the symmetric case is slightly better.
Evaluating the efficiencies with samples where the charge on all strips is scaled by \SI{10}{\percent} up or down to account for the observed mismodelling in the simulation leads to a relative change of at most \SI{4}{\percent} in the signal efficiency and \SI{3.5}{\percent} in the background rejection.

\autoref{fig:signal-efficiencies} shows the efficiencies for di-photon separations of $d < \SI{300}{\micro\meter}$ and $\SI{500}{\micro\meter} < d < \SI{1000}{\micro\meter}$ for a working point with \SI{90}{\percent} average background rejection. As expected, it can be seen that the efficiency for small photon separations is lower compared to larger separations. In fact, the signal efficiency ranges from \SI{19}{\percent} to \SI{79.5}{\percent} for two photons separated by \SI{100}{\micro\meter} to \SI{2000}{\micro\meter}, respectively, with energies above \SI{100}{\GeV}, each, at an average background rejection rate of \SI{90}{\percent} for a single photon with an energy above \SI{200}{\GeV}.

\section{Applications in Axion Search Experiments}

As already previously motivated, the developed pre-shower detector could be directly applied in future Axion search experiments, such as FASER. Hence we estimate in the following the expected performance of the detector system for a concrete example. The kinematics of di-photons from ALP decays depend on the axion mass, $m_a$, the initial axion momentum as well as the coupling to photons $c_{a\gamma\gamma}$. The smaller the axion mass and the higher the axion momentum, the smaller the expected average distance of the two photons. Moreover, the energy of the two photons can be highly asymmetric depending on the relative orientation of the axion-decay in its restframe to the axion boost direction. A symmetric energy distribution between the two decay photons is therefore only expected if the decay is perpendicular to the axion momentum. 

The physics reach for the observation of an axion signal with a confidence level of $3\sigma$ at the FASER experiment for several sencarios has been studied. The expected axion signal events are taken from simulation and are evaluated for 1452 different combinations of the ALP mass $m_a$ and coupling $g_{aWW}$ in the range $m_a = \SIrange[range-phrase={\text{ -- }}]{0.1}{2}{\GeV}$ and $g_{aWW} = \SIrange[range-phrase={\text{ -- }}]{5e-7}{8e-4}{\per\GeV}$.
The integrated luminosity is fixed to \SI{90}{\per\femto\barn} reflecting a fraction of LHC-Run 2 and \SI{3}{\per\atto\barn} reflecting the High-Luminosity LHC.
As a detailed simulation of the background is not available at point, the reach is evaluated for the assumptions of 10 or 200 background events for \SI{90}{\per\femto\barn} and 333 or 6666 background events for \SI{3}{\per\atto\barn}, respectively.

The number of observed events is calculated by convoluting the number of expected events with a given kinematic in dependence of the two photon energies and their distance with the corresponding expected signal efficiencies of our pre-shower detector. 
The resulting physics yield is summarized in \autoref{fig:physics-reach} and has been evaluated for two different working points with a background rejection of \SI{90}{\percent} (loose) and \SI{99.5}{\percent} (tight), with corresponding average signal efficiencies of approximately \SI{45}{\percent} and \SI{30}{\percent}.
The systematic variation of the measured charge has negligible influence on the physics yield. The tight classification option significantly outperforms the loose option. Moreover it becomes evident that the pre-shower detectors extends the physics reach by a factor of three in axion mass and and a factor of five in the axion coupling strength.

\section{Conclusion}

A preshower detector based on three active layers of Micromegas detectors was presented, which is able to separate high energetic one-photon from two-photon events. Each of those Micromegas detectors has a depth of \SI{1.5}{\centi\meter}, is spark-resistant due to a resistive protection layer and allows for a spatial resolution of below \SI{120}{\micro\meter}, even at very high rates. The measured signals of the active layers are fed to a deep neural network, which acts as binary classifier.
The efficiency ranges from \SI{19}{\percent} to \SI{79.5}{\percent} for two high energetic photons separated by \SI{100}{\micro\meter} to \SI{2000}{\micro\meter}, respectively with a background rejection rate of \SI{90}{\percent} for single photons. The efficiency is thereby rising fast with increasing photon distance reaching the plateau at \SI{1000}{\micro\meter}. The energy dependence is relatively small and the efficiency only drops significantly for one of the photons being less than \SI{200}{\GeV}. 
In a second step, we estimated the increase of the sensitivity to the search for axion like particles at the FASER experiment when using the pre-shower detector. It was found that an increase by several factors in axion mass and coupling can be expected. In summary, the developed pre-shower detector design as well as the prototype detector offer a cost effective solution for the event classification of future experiments that aim for the discovery of ALPs in a mass region between \SI{1}{\MeV} and \SI{5}{GeV} in the di-photon final state.

\section*{Acknowledgement}
We would like to thank the MAMI accelerator team, in particular P. Guelker and W. Lauth for their help during our test beam measurements. We also would thank Didier Ferrere and Giuseppe Iacobucci for their help in the context of general considerations of the development of pre-shower detectors in the context of axion searches. This work is supported by the European Commission for Research (ERC) in the context of the ERC Consolidator Grant LightAtTheLHC.

%\bibliographystyle{apsrev4-1} %%% physical review (up to date)
%\bibliography{./Bibliography.bib}
\printbibliography%

@article{APVJones,
  title = {The {{APV25}} Deep Sub {{Micron}} Readout Chip for {{CMS}} Channels Detectors},
  author = {{L.Jones} and others},
  date = {1999},
  journaltitle = {Proceedings of 5th workshop on Chips electronics for LHC experiments},
  volume = {CERN/LHCC/99-09},
  pages = {162--166},
  doi = {10.1088/1748-0221/8/03/C03015}
}

@article{Lin:2014jxa,
    author = {Lin, Tai-Hua and D\"udder, Andreas and Schott, Matthias and Valderanis, Chrysostomos and Wehner, Laura and Westenberger, Robert},
    title = "{Signal Characteristics of a Resistive-Strip Micromegas Detector with an Integrated Two-Dimensional Readout}",
    eprint = "1406.6871",
    archivePrefix = "arXiv",
    primaryClass = "physics.ins-det",
    doi = "10.1016/j.nima.2014.09.002",
    journal = "Nucl. Instrum. Meth. A",
    volume = "767",
    pages = "281--288",
    year = "2014"
}

@article{Bauer:2017nlg,
    author = "Bauer, Martin and Neubert, Matthias and Thamm, Andrea",
    title = "{LHC as an Axion Factory: Probing an Axion Explanation for $(g-2)_\mu$ with Exotic Higgs Decays}",
    eprint = "1704.08207",
    archivePrefix = "arXiv",
    primaryClass = "hep-ph",
    doi = "10.1103/PhysRevLett.119.031802",
    journal = "Phys. Rev. Lett.",
    volume = "119",
    number = "3",
    pages = "031802",
    year = "2017"
}

@article{Byszewski:2012zz,
  title = {Resistive-Strips Micromegas Detectors with Two-Dimensional Readout},
  author = {Byszewski, M. and Wotschack, J.},
  date = {2012},
  journaltitle = {JINST},
  volume = {7},
  pages = {C02060},
  doi = {10.1088/1748-0221/7/02/C02060}
}

@online{faserOverview,
  title = {{{FASER}}: {{ForwArd Search ExpeRiment}} at the {{LHC}}},
  shorttitle = {{{FASER}}},
  author = {{FASER Collaboration}},
  date = {2019-01},
  eprint = {1901.04468},
  eprinttype = {arxiv},
  primaryclass = {hep-ex, physics:hep-ph, physics:physics},
  url = {http://arxiv.org/abs/1901.04468},
  archiveprefix = {arXiv}
}

@article{faserPhysicsReach,
  title = {{{FASER}}'s {{Physics Reach}} for {{Long}}-{{Lived Particles}}},
  author = {{FASER Collaboration}},
  date = {2019-05},
  journaltitle = {Physical Review D},
  volume = {99},
  number = {9},
  pages = {095011},
  issn = {2470-0010, 2470-0029},
  doi = {10.1103/PhysRevD.99.095011},
  url = {http://arxiv.org/abs/1811.12522}
}

@article{geant4,
  title = {Geant4—a Simulation Toolkit},
  author = {Agostinelli, S. and Allison, J. and Amako, K. and Apostolakis, J. and Araujo, H. and Arce, P. and Asai, M. and Axen, D. and Banerjee, S. and Barrand, G. and Behner, F. and Bellagamba, L. and Boudreau, J. and Broglia, L. and Brunengo, A. and Burkhardt, H. and Chauvie, S. and Chuma, J. and Chytracek, R. and Cooperman, G. and Cosmo, G. and Degtyarenko, P. and Dell'Acqua, A. and Depaola, G. and Dietrich, D. and Enami, R. and Feliciello, A. and Ferguson, C. and Fesefeldt, H. and Folger, G. and Foppiano, F. and Forti, A. and Garelli, S. and Giani, S. and Giannitrapani, R. and Gibin, D. and Gómez Cadenas, J.J. and González, I. and Gracia Abril, G. and Greeniaus, G. and Greiner, W. and Grichine, V. and Grossheim, A. and Guatelli, S. and Gumplinger, P. and Hamatsu, R. and Hashimoto, K. and Hasui, H. and Heikkinen, A. and Howard, A. and Ivanchenko, V. and Johnson, A. and Jones, F.W. and Kallenbach, J. and Kanaya, N. and Kawabata, M. and Kawabata, Y. and Kawaguti, M. and Kelner, S. and Kent, P. and Kimura, A. and Kodama, T. and Kokoulin, R. and Kossov, M. and Kurashige, H. and Lamanna, E. and Lampén, T. and Lara, V. and Lefebure, V. and Lei, F. and Liendl, M. and Lockman, W. and Longo, F. and Magni, S. and Maire, M. and Medernach, E. and Minamimoto, K. and Mora de Freitas, P. and Morita, Y. and Murakami, K. and Nagamatu, M. and Nartallo, R. and Nieminen, P. and Nishimura, T. and Ohtsubo, K. and Okamura, M. and O'Neale, S. and Oohata, Y. and Paech, K. and Perl, J. and Pfeiffer, A. and Pia, M.G. and Ranjard, F. and Rybin, A. and Sadilov, S. and Di Salvo, E. and Santin, G. and Sasaki, T. and Savvas, N. and Sawada, Y. and Scherer, S. and Sei, S. and Sirotenko, V. and Smith, D. and Starkov, N. and Stoecker, H. and Sulkimo, J. and Takahata, M. and Tanaka, S. and Tcherniaev, E. and Safai Tehrani, E. and Tropeano, M. and Truscott, P. and Uno, H. and Urban, L. and Urban, P. and Verderi, M. and Walkden, A. and Wander, W. and Weber, H. and Wellisch, J.P. and Wenaus, T. and Williams, D.C. and Wright, D. and Yamada, T. and Yoshida, H. and Zschiesche, D.},
  date = {2003-07},
  journaltitle = {Nuclear Instruments and Methods in Physics Research Section A: Accelerators, Spectrometers, Detectors and Associated Equipment},
  volume = {506},
  number = {3},
  pages = {250--303},
  issn = {01689002},
  doi = {10.1016/S0168-9002(03)01368-8},
  url = {https://linkinghub.elsevier.com/retrieve/pii/S0168900203013688},
  langid = {english}
}

@misc{keras,
  title = {Keras},
  author = {Chollet, François and others},
  date = {2015},
  url = {https://keras.io}
}

@inproceedings{Martoiu:2011zja,
  title = {Front-End Electronics for the {{Scalable Readout System}} of {{RD51}}},
  booktitle = {2011 {{IEEE Nuclear Science Symposium Conference Record}}},
  author = {Martoiu, S. and Muller, H. and Toledo, J.},
  date = {2011-10},
  pages = {2036--2038},
  publisher = {{IEEE}},
  location = {{Valencia, Spain}},
  doi = {10.1109/NSSMIC.2011.6154414},
  url = {http://ieeexplore.ieee.org/document/6154414/},
  eventtitle = {2011 {{IEEE Nuclear Science Symposium}} and {{Medical Imaging Conference}} (2011 {{NSS}}/{{MIC}})},
  isbn = {978-1-4673-0120-6}
}

@article{Martoiu:2013aca,
  title = {Development of the Scalable Readout System for Micro-Pattern Gas Detectors and Other Applications},
  author = {Martoiu, S and Muller, H and Tarazona, A and Toledo, J},
  date = {2013-03-12},
  journaltitle = {J. Inst.},
  volume = {8},
  number = {03},
  pages = {C03015-C03015},
  issn = {1748-0221},
  doi = {10.1088/1748-0221/8/03/C03015},
  url = {https://iopscience.iop.org/article/10.1088/1748-0221/8/03/C03015}
}

@article{tensorflow,
  title = {{TensorFlow}: Large-Scale Machine Learning on Heterogeneous Systems},
  author = {Abadi, Martín and Agarwal, Ashish and Barham, Paul and Brevdo, Eugene and Chen, Zhifeng and Citro, Craig and Corrado, Greg S. and Davis, Andy and Dean, Jeffrey and Devin, Matthieu and Ghemawat, Sanjay and Goodfellow, Ian and Harp, Andrew and Irving, Geoffrey and Isard, Michael and Jia, Yangqing and Jozefowicz, Rafal and Kaiser, Lukasz and Kudlur, Manjunath and Levenberg, Josh and Mané, Dandelion and Monga, Rajat and Moore, Sherry and Murray, Derek and Olah, Chris and Schuster, Mike and Shlens, Jonathon and Steiner, Benoit and Sutskever, Ilya and Talwar, Kunal and Tucker, Paul and Vanhoucke, Vincent and Vasudevan, Vijay and Viégas, Fernanda and Vinyals, Oriol and Warden, Pete and Wattenberg, Martin and Wicke, Martin and Yu, Yuan and Zheng, Xiaoqiang},
  date = {2015},
  url = {https://www.tensorflow.org/}
}

@article{klingParameterSpace,
  title = {Looking Forward to Test the {{KOTO}} Anomaly with {{FASER}}},
  author = {Kling, Felix and Trojanowski, Sebastian},
  date = {2020-07-31},
  journaltitle = {Phys. Rev. D},
  volume = {102},
  number = {1},
  eprint = {2006.10630},
  eprinttype = {arxiv},
  pages = {015032},
  issn = {2470-0010, 2470-0029},
  doi = {10.1103/PhysRevD.102.015032},
  url = {http://arxiv.org/abs/2006.10630},
  urldate = {2021-10-07},
  archiveprefix = {arXiv}
}

@article{CPProblem1,
  title = {CP Conservation in the Presence of pseudoparticles},
  author = {R. D. Peccei and H. R. Quinn},
  date = {1977},
  journaltitle = {Phys.Rev.Lett.},
  volume = {38(25)},
  pages = {1440},
  doi = {10.1103/PhysRevLett.38.1440}
}

@article{CPProblem2,
  title = {Constraints imposed by CP conservation in the presence of pseudoparticles},
  author = {R. D. Peccei and H. R. Quinn},
  date = {1977},
  journaltitle = {Phys.Rev.D},
  volume = {16},
  pages = {1791},
  doi = {10.1103/PhysRevD.16.1791}
}

@article{Axions1,
  title = {A New Light Boson?},
  author = {S. Weinberg},
  date = {1978},
  journaltitle = {Phys.Rev.Lett.},
  volume = {40},
  pages = {223},
  doi = {10.1103/PhysRevLett.40.223}
}

@article{Axions2,
  title = {Problem of Strong P and T Invariance in the Presence of Instantons},
  author = {F. Wilczek},
  date = {1978},
  journaltitle = {Phys.Rev.Lett.},
  volume = {40},
  pages = {279},
  doi = {10.1103/PhysRevLett.40.279}
}

@article{AxionsDM,
  title = {WISPy Cold Dark Matter},
  author = {M. Goodsell J. Jaeckel J. Redondo P. Arias, D. Cadamuro and A. Ringwald.},
  date = {2012},
  journaltitle = {JCAP},
  volume = {2012},
  number = {06},
  pages = {013},
  doi = {10.1088/1475-7516/2012/06/013}
}

\end{document}